# Controlling Short-channel Effects in Deep Submicron SOI MOSFETs for Improved Reliability: A Review

Anurag Chaudhry and M. Jagadesh Kumar[1]

Department of Electrical Engineering,

Indian Institute of Technology, Delhi,

Hauz Khas, New Delhi – 110 016, INDIA.

Email: mamidala@ieee.org Fax: 91-11-2658 1264

---

[1]Corresponding author

## Abstract

This paper examines the performance degradation of a MOS device fabricated on silicon-on-insulator (SOI) due to the undesirable short-channel effects (SCE) as the channel length is scaled to meet the increasing demand for high-speed high-performing ULSI applications. The review assesses recent proposals to circumvent the SCE in SOI MOSFETs and a short evaluation of strengths and weaknesses specific to each attempt is presented. A new device structure called the *dual-material gate* (DMG) SOI MOSFET is discussed and its efficacy in suppressing SCEs such as *drain-induced barrier lowering* (DIBL), *channel length modulation* (CLM) and hot-carrier effects, all of which affect the reliability of ultra-small geometry MOSFETs, is assessed.

# 1. Introduction

In order to realize higher-speed and higher-packing density MOS integrated circuits, the dimensions of MOSFET's have continued to shrink according to the scaling law proposed by Dennard *et al.* [1]. Yet, the power consumption of modern VLSI's has become rather significant as a result of extremely large integration. Reducing this power is strongly desired. Choosing a lower power supply voltage is an effective method. However, it leads to the degradation of MOSFET current driving capability. Consequently, scaling of MOS dimensions is important in order to improve the drivability, and to achieve higher-performance and higher-functional VLSI's.

With aggressive technology scaling to enhance performance, circumventing the detrimental short-channel effects (SCE) to improve the device reliability has been the focus in MOSFET scaling. When the channel length shrinks, the controllability of the gate over the channel depletion region reduces due to the increased charge sharing from source/drain. SCE leads to several reliability issues including the dependence of device characteristics, such as threshold voltage, upon channel length. This leads to the scatter of device characteristics because of the scatter of gate length produced during the fabrication process. The predominating reliability problems associated with SCE are a lack of pinchoff and a shift in threshold voltage with decreasing channel length as well as drain induced barrier lowering (DIBL) and hot-carrier effect at increasing drain voltage. Moreover, SCE degrades the controllability of the gate voltage to drain current, which leads to the degradation of the subthreshold slope and the increase in drain off-current. This degradation is described as charge sharing by the gate and drain electric fields in the channel depletion layer in Poon and Yau's model [2], which was reported as the first SCE model. Thinning gate oxide and using shallow source/drain junctions are known to be effective ways of preventing SCE. With short-channel devices, the reliability margins have also been cut down significantly [3]. Particularly, the high electric field near the drain becomes more crucial and poses a limit on device operation, notably by a large gate current, substrate current and a substantial threshold voltage shift [4-9]. Efforts have been made to model the device degradation due to hot electron generation [10-16].

This description can be applied to conventional MOSFET's fabricated in a bulk silicon wafer. What about thin-film SOI MOSFET's ? They are attractive devices for low-power high-speed VLSI applications because of their small parasitic capacitance [17]. Young [18] analyzed the SCE using a device simulator, and concluded that SCE is well suppressed in thin-film SOI MOSFET's when compared to bulk MOSFET's. In general, it is believed that thin-film SOI MOSFET's have a higher immunity to SCE compared with bulk MOSFET's. This may be due to the difference in source/drain junction depths between the two kinds of devices. For instance, the thickness of the silicon film, $t_{Si}$, which corresponds to the source/drain junction depth of 50–100 nm, is common in 0.25–0.35 μm SOI MOSFET's. It is extremely shallow compared with the junction depth of 100–200 nm in 0.25–0.35 μm gate bulk MOSFET's. However, to take advantage of the ameliorated SCEs in deep-submicron fully-depleted SOI, $t_{Si}$ must be considerably smaller than the source/drain junction depth ($t_{Si} \sim$ 10-15 nm). Moreover, a strong coupling through the buried oxide in thin-film devices exists and consequently, very thin buried oxides ($t_b \sim$ 100 nm) are needed which trade-offs with junction capacitance considerations. With the gate length scaling approaching sub-100-nm regime for improved performance and density, the requirements for body-doping concentration, gate oxide thickness, and source/drain (S/D) doping profiles to control short-channel effects become increasingly difficult to meet when conventional device structures based on bulk silicon substrates are employed. Moreover SOI brings in new reliability issues, which are not known in the traditional bulk-Si devices, related to the presence of the buried oxide like self-heating and hot-electron degradation of the buried oxide. In a high electrical field of a short-channel transistor, carriers may gain enough energy and get trapped in the buried oxide along with the gate oxide. The buried oxide is more subject to degradation than the gate oxide because the high density of electron traps is an intrinsic feature of SIMOX oxides. These defects may change parameters of the back channel in FET and affect the performance of CMOS circuit through the coupling effect. Thus, the hot carrier induced degradation in SOI devices is more complex than that in bulk devices because of the thin-film effects and the existence of two interfaces (two oxides and two channels). Hence, for small-geometry SOI CMOS devices, short-channel effects are becoming increasingly important [19]-[28]

Several novel device structures have been reported in literature to circumvent the undesirable SCE in SOI devices. In this paper an understanding of the physical mechanisms determining SCE is espoused and the recent proposals are reviewed and assessed in light of their approach to mitigate the SCE. A new device structure called the *dual-material gate* (DMG) is introduced and its efficacy in suppressing SCE in thin-film SOI MOSFETs is highlighted using an analytical model which incorporates the effect of device parameters like source/drain and body doping concentrations, the lengths of the gate metals and their work functions, applied drain and substrate biases, the thickness of the gate and buried oxide.

# 1. Short-channel effects in SOI MOS Devices

Short-channel effects (SCE) can be chiefly attributed to the so-called drain-induced barrier lowering (DIBL) effect which causes a reduction in the threshold voltage as the channel length decreases. But, in an SOI device SCE is also influenced by thin-film thickness, thin-film doping density, substrate biasing and buried oxide thickness.

## 2.2. *Drain-Induced Barrier Lowering (DIBL)*

In the weak inversion regime there is a potential barrier between the source and the channel region. The height of this barrier is a result of the balance between drift and diffusion current between these two regions. The barrier height for channel carriers should ideally be controlled by the gate voltage to maximize transconductance. As indicated in Fig. 1, *drain-induced barrier lowering* (DIBL) effect [29] occurs when the barrier height for channel carriers at the edge of the source reduces due to the influence of drain electric field, upon application of a high drain voltage. This increases the number of carriers injected into the channel from the source leading to an increased drain off-current. Thus the drain current is controlled not only by the gate voltage, but also by the drain voltage.

For device modeling purposes this parasitic effect can be accounted for by a threshold voltage reduction depending on the drain voltage [30].

In addition to the surface DIBL, there are two unique features determining SCEs in thin-film SOI devices *viz.* (a) positive bias effect to the body due to the accumulation of holes generated by impact ionization near the drain and (b) the DIBL effect on the barrier height for holes at the edge of the source near the bottom of thin-film, as illustrated in Fig. 2 [31].

Holes generated near the drain due to impact ionization accumulate in the body region, and then positively bias the body, reducing threshold voltage, $V_{th}$. This positive bias effect leads to $V_{th}$ lowering for all gate lengths, including rather long gates such as 2 µm. The hole generation rate due to impact ionization increases as gate length decreases under a fixed value of $V_{DS}$. This effect is predominant in PD SOI nMOSFETs and results in so-called *floating body effects* (FBE) [32], [33].

The DIBL effect on the barrier height for holes reduces the positive bias effect to the body because the accumulated holes in the body can more easily surmount the barrier and flow to the source. As a result fewer number of accumulated holes remain which weakens the $V_{th}$ lowering. The potential near the bottom in the body region increases as gate length decreases due to the drain electric field. This leads to the lowering of the barrier height for holes at the source edge near the bottom with shorter gate lengths. Fig. 3 compares the schematic energy band diagrams at threshold condition between short and long channels MOSFET's. The comparison is done near the bottom of the thin-film from the source to the drain. With shorter gate lengths, the barrier height for holes near the bottom is lowered by the influence of the drain electric field, and holes accumulated in the body region can more easily flow into the source.

Due to these three mechanisms, $V_{th}$ dependence upon gate length in FD nMOSFET's becomes small, as illustrated in Fig. 4.

## 2.2. Back-Gate Biasing dependence

Fig. 5 shows the short-channel effect of the FD SOI NMOS device with a front gate oxide of 9.2 nm, a buried oxide of 400 nm, and a thin-film of 80 nm, biased at the back gate bias of 0 and –5 V [34].

As shown in the Fig. 5, at a negative back gate bias of –5 V, the threshold voltage is lifted upward as compared to the back gate bias of 0 V. The extent of the upward shift when the back gate bias becomes negative is smaller for a device with shorter channel length, which implies that SCE seems to improve. With a shorter channel, the controllability over the vertical direction of the channel region from the source/drain seems to be reduced at a more negative back gate bias, hence its back gate bias effect is smaller.

### 2.3. *Structure dependence*

In addition to the drain and back gate biasing dependences, the SCE of an SOI MOS device is also influenced by the thin-film thickness. Fig. 6 shows the threshold voltage roll-off of the FD SOI NMOS device with a front gate oxide of 4.5 nm for various thin-film thicknesses [35].

As shown in Fig. 6, when the thin-film thickness is reduced, for both NMOS and PMOS devices, the SCE becomes smaller since the controllability of the front gate over the active channel region is stronger and the source/drain has less influence in the channel.

The short channel effect is also dependent on the thin-film doping density. Fig. 7 shows the threshold voltage shift versus the thin-film thickness of an SOI NMOS device with a front gate oxide of 5 nm and a buried oxide of 360 nm for various channel doping densities, biased at (a) $V_{DS} = 0.05$ V, and (b) 1.5 V [36]. As shown in Fig. 7, when the thin-film thickness exceeds a critical thickness the device operates in the PD regime. Below this specific thickness the device operates in the FD regime. In the FD regime, SCE is smaller with a lighter thin-film doping density, which is opposite to that in the PD regime.

The influence of source/drain to the channel region via the buried oxide can also worsen the SCE. Fig. 8 shows the short-channel effect of an SOI NMOS device with a front gate oxide of 6 nm, thin-film of 100 nm for buried oxide thickness of 100 nm and 400 nm [37]. For a device with thinner buried oxide, the SCE is lessened. With thinner buried oxide, the compressive stress is higher. Hence, during the thermal process in fabrication, boron dopants in the thin film cannot diffuse easily. As a result, the doping density of thin-film is higher and its threshold voltage is higher. As the doping density of thin-film is raised, the SCE is reduced.

**2. Solutions to contain Short-channel Effects**

One of the primary reason for device degradation at shorter channel lengths is the encroachment of drain electric field in the channel region as shown in Fig. 9. As shown in the figure, the gate electrode shields the channel region from those lines at the top of the device, but electric field lines penetrate the device laterally and from underneath, through the buried oxide and the silicon wafer substrate causing the undesirable DIBL for the charge carriers.

Several device structures have been proposed to alleviate the degrading effect of the drain electric field on device performance of sub-micron SOI MOSFET's as discussed below.

*3.2    Thin body FD SOI with raised source and drain*

Reduction of short-channel effects in FD SOI MOSFETs requires the use of thin silicon films to eliminate the sub-surface leakage paths. A device structure that implements this concept is the thin-body MOSFET [39]-[40]. In thin-body MOSFET, the source-to-drain current is restricted to flow in a region close to the gate for superior gate control, as illustrated in Fig. 10. Since it does not rely on a heavily-doped channel for the suppression of short-channel effects, it avoids the problems of mobility degradation due to impurity scattering and threshold voltage fluctuation due to the random variation of the number of dopant atoms in the channel region of nanoscale transistors [41].

The device shown in Fig. 10 has a thin-body on insulator structure [43], [44] and is essentially an extension of the fully depleted SOI transistor. Since a thin source/drain (S/D) region would contribute a high series resistance that degrades the drive current, a raised S/D is introduced to avoid the series resistance problem. Reference [44] demonstrated raised S/D formation by poly-Si deposition followed by an etch-back. Nevertheless, parasitic capacitances between the raised S/D and the gate are inherent in this device structure. This is expected to adversely impact the device speed and power consumption. An attempt to reduce the parasitic capacitance by increasing the distance between the raised S/D and the gate leads to an increase in series resistance.

### 3.2   *Metal Source and Drain FDSOI MOSFET*

Another proposed technique for reducing the source and drain resistance in thin-film FDSOI MOSFETs consists in using metal (or silicide) source and drain. However, the formation of Schottky barriers between the source/drain and the channel must be avoided. The formation of a low (ideally zero) Schottky barrier is needed to insure the formation of an ohmic contact between the source/drain and the channel. Since the Schottky barrier varies with the applied gate bias in inversion-mode devices, it is more appropriate to use accumulation-mode devices when metal source/drain structures are used, as the surface potential remains constant when an accumulation channel is created

[45][46].

### 3.3 Metal gate FDSOI

As the transistors are aggressively scaled down to sub-100 nm, problems such as poly-Si gate depletion, boron penetration, and high gate resistance are aggravated [47]. Alternative gate electrodes, such as metal gates, are promising to address these issues. Fig. 11 shows the threshold voltage versus the channel length of an FD SOI NMOS device with a front gate oxide of 5 nm, a thin film of 100 nm and a buried oxide of 420 nm, using polysilicon and tantalum gates [48]. The use of tantalum gate is to facilitate the adjustment of the threshold voltage of an SOI device without raising the thin-film doping density substantially by taking advantage of the workfunction of tantalum. By using metal (tantalum) as the front-gate material, the problem of polysilicon gate depletion associated with polysilicon gates is removed and therefore, SCE is smaller.

For PD SOI, metal gates with workfunction of 0.1 ~ 0.2 eV away from the silicon band edges enable the use of relatively low halo dose. This reduces the possibility of band-to-band tunneling without compromising performance. Whereas, for an FD SOI, a metal gate with workfunction close to the band edges would require a high channel doping to meet the off-current specifications. The need for high doping concentration increases $V_{th}$ fluctuations due to variation in thin-film thickness in addition to serious mobility degradation. Midgap gates are desirable for FD SOI MOSFETs in such a scenario [49].

### 3.4 Buried Insulator engineering

Fig. 12 shows the variation of threshold voltage roll-off due to DIBL and charge sharing (CS) with permittivity of buried oxide for SOI MOSFETs with channel lengths

30 nm and 500 nm [50]. The reduction of buried oxide permittivity improves the DIBL effect due to the reduced field penetration into the buried oxide from the drain, but, it does not affect the charge sharing significantly.

### 3.5 Graded Channel FDSOI

Fig. 13 shows the threshold voltage versus channel length of an FD SOI NMOS device with a front gate oxide of 7 nm, a thin-film of 50 nm and a buried oxide of 120 nm for a (a) uniformly doped channel and (b) graded channel [51]. In the device with graded channel, in the centre of the channel, the doping density is the same as for the device with uniformly doped channel whereas near source/drain regions more highly doped regions are generated via the gate-edge (GE) implant techniques. As shown in the Fig.13, compared to the uniformly doped channel, GE implanted graded channel improves the SCE substantially, especially at large drain voltage.

Increasing the doping density of the thin-film can reduce short-channel effects in PD SOI devices. However, a very high doping density of the thin-film may lead to an undesirable excessive magnitude in the threshold voltage.

### 3.6 HALO Doped SOI

With continuous device scaling down to 100 nm channel length and less, the HALO (or pocket) implantations have been introduced to better control the short-channel effects. In digital applications HALO implantations have the purpose of reducing the off-state leakage current while maximizing transistor linear and saturated drive currents. While for analog applications it has been shown that HALO implantation is needed for base-band applications using longer channel, it has detrimental effect for high speed applications using minimum channel transistors in strong inversion [52]. Excessive

HALO implantation in PD SOI transistors increases the kink effect. HALO implantation is also known to degrade the distortion characteristics when the SOI devices are used as resistors [52]. Taur [53] demonstrated that a *super-halo*, a highly non-uniform 2-D dopant profile in the channel and the body region effectively controls short-channel effects in 25 nm MOSFET. A properly scaled *super-halo* is able to suppress the potential barrier lowering both in the inversion and the body depletion region. When strong halo is used, drain-halo (or body) band-to-bad tunneling leakage can be a considerable contributor to the total off-state leakage current at room temperature. Substrate-injection gate current also increases in devices with stronger halo implant.

Recently, asymmetric single halo (SH) MOSFET structures have been introduced for bulk [54-55] as well as for SOI MOSFETs [56-57] to adjust the threshold voltage and improve the device SCE and hot carrier effects (HCE). These devices also achieve higher drive currents by exploiting the velocity overshoot phenomenon [54], which is an advantage in mixed mode analog/digital circuits. The schematic cross section of a typical SH SOI n-type MOSFET is shown in Fig.14 [58]. It has been shown that these devices show a marginal improvement in transconductance and lower output conductance as compared to the conventional SOI devices. The other advantages of SH devices over conventional SOI like absence of kink, lower inherent parasitic bipolar junction transistor (pBJT) gain have also been reported [59-60].

### 3.2    *Ground-Plane FDSOI MOSFET*

To keep electric field lines from the drain from propagating into the channel region a ground-plane can be formed in the silicon substrate underneath the buried oxide. Fig. 15 shows that a heavily doped electric-field stop can be placed in the substrate either underneath the boundary between channel and source/drain or underneath the channel region itself. This field stop effectively improves SCE and subthreshold slope [61][62].

*3.3   Multiple-Gate FDSOI MOSFET*

To prevent the encroachment of electric field lines from the drain on the channel region, special gate structures can be used as shown in Fig. 16. Such "multiple"-gate devices include double-gate transistors, triple-gate devices such as the quantum wire [65], the FinFET [66] and Δ-channel SOI MOSFET [67], and quadruple-gate devices such as the gate-all-around device [21], the DELTA transistor [68][69], and vertical pillar MOSFETs [70],[71].

The double-gate concept was first reported in 1984 [73] and has been fabricated by several groups since then. The use of a double gate results in enhanced transconductance, due to the volume inversion effect [22][74] and better subthreshold slope. The fabrication process, however, is considered unpractical for commercial applications because it uses lateral epitaxial overgrowth or the etching of a cavity underneath the devices [21][75]. Also, since the thickness of silicon between the two gates is smaller than the physical gate length, the most critical lithography step in printing the double-gate transistor becomes patterning of the thin-film, rather than the physical gate length patterning [76].

Fig. 17 shows the DIBL and threshold voltage roll-off as a function of gate voltage for double, triple, quadruple and Π-gate devices. The best performance is obtained from the quadruple gate, but Π-gate is close second. The results show the efficient shielding of the channel by the gate electrode from the electric field lines originating from the drain region.

## 3. Dual-Material Gate structure

Dual-Material Gate (DMG) structure employs "gate-material engineering" instead of "doping engineering" with different workfunctions to introduce a potential step in the channel [77]. This leads to a suppression of short-channel effects and an enhanced source side electric field resulting in increased carrier transport efficiency in the channel region. A schematic cross-sectional view of a DMG fully depleted SOI MOSFET is shown in Fig. 18 with gate metals M1 and M2 of lengths $L_1$ and $L_2$, respectively. In a n-channel DMG SOI MOSFET, the work function of metal gate 1 (M1) is greater than metal gate 2 (M2) i.e., $\phi_{M1} > \phi_{M2}$ and vice-versa for a p-channel MOSFET.

A physics based 2-D model for the surface potential variation along the channel can be developed by solving the two-dimensional Poisson's equation to analyze the SCE suppression achieved with a fully depleted Dual-Material Gate (DMG) silicon-on-insulator MOSFET's [78]. Numerical simulations are used to validate this model predictions and compare the performance of thin-film DMG SOI with the single-material gate (SMG) SOI MOSFETs.

### *4.2 Mathematical Formulation*

Assuming that the impurity density in the channel region is uniform and the influence of charge carriers on the electrostatics of the channel can be neglected, the potential distribution in the silicon thin-film, before the onset of strong inversion can be written as

$$\frac{d^2\phi(x,y)}{dx^2} + \frac{d^2\phi(x,y)}{dy^2} = \frac{qN_A}{\varepsilon_{Si}} \quad \text{for } 0 \leq x \leq L, \ 0 \leq y \leq t_{Si} \quad (1)$$

where $N_A$ is the film doping concentration, $\varepsilon_{Si}$ is the dielectric constant of silicon, $t_{Si}$ is the film thickness and $L$ is the device channel length. The potential profile in the vertical direction, i.e., the $y$-dependence of $\phi(x,y)$ can be approximated by a simple parabolic function as proposed by Young [18] for fully depleted SOI MOSFET's.

$$\phi(x,y) = \phi_S(x) + c_1(x)y + c_2(x)y^2 \qquad (2)$$

where $\phi_S(x)$ is the surface potential and the arbitrary coefficients $c_1(x)$ and $c_2(x)$ are functions of $x$ only. In the DMG structure, since the gate is divided into two parts the potential under M1 and M2 can be written as

$$\phi_1(x,y) = \phi_{S1}(x) + c_{11}(x)y + c_{12}(x)y^2 \quad \text{for } 0 \leq x \leq L_1, \ 0 \leq y \leq t_{Si} \qquad (3)$$

$$\phi_2(x,y) = \phi_{S2}(x) + c_{21}(x)y + c_{22}(x)y^2 \quad \text{for } L_1 \leq x \leq L_1 + L_2, \ 0 \leq y \leq t_{Si} \qquad (4)$$

The Poisson's equation is solved separately under the two gate regions using the boundary conditions stated in [78] to obtain the following solution:

$$\phi_{S1}(x) = A\exp(\lambda_1 x) + B\exp(\lambda_2 x) - \frac{\beta_1}{\alpha} \qquad (5)$$

$$\phi_{S2}(x) = C\exp(\lambda_1(x - L_1)) + D\exp(\lambda_2(x - L_1)) - \frac{\beta_2}{\alpha} \qquad (6)$$

where the constants $\beta_1$, $\beta_2$ and $\alpha$ are as defined in [78]. The coefficients $A$, $B$, $C$ and $D$ of the exponent in the above equations are

$$A = \left\{ \frac{(V_{bi} - \sigma_2 + V_{DS}) - \exp(-\lambda_1(L_1 + L_2))(V_{bi} - \sigma_1) - (\sigma_1 - \sigma_2)\cosh(\lambda_1 L_2)}{1 - \exp(-2\lambda_1(L_1 + L_2))} \right\} \exp(-\lambda_1(L_1 + L_2))$$

$$B = \frac{(V_{bi} - \sigma_1) - (V_{bi} - \sigma_2 + V_{DS})\exp(-\lambda_1(L_1 + L_2)) + (\sigma_1 - \sigma_2)\cosh(\lambda_1 L_2)\exp(-\lambda_1(L_1 + L_2))}{1 - \exp(-2\lambda_1(L_1 + L_2))}$$

$$C = A\exp(\lambda_1 L_1) + \frac{(\sigma_1 - \sigma_2)}{2} \qquad \text{and} \qquad D = B\exp(\lambda_2 L_1) + \frac{(\sigma_1 - \sigma_2)}{2}$$

where $\sigma_1 = -\beta_1/\alpha$, $\sigma_2 = -\beta_2/\alpha$.

The minimum potential of the front-channel can be calculated from (5) as

$$\phi_{S1,min} = 2\sqrt{AB} + \sigma_1 \tag{7}$$

The minima occurs at

$$x_{min} = \frac{1}{2\lambda_1}\ln\left(\frac{B}{A}\right) \tag{8}$$

### *4.3 Results and Discussion*

DIBL effect can be demonstrated by plotting the surface potential minima $\phi_{S1,min}$, as a function of the position along the channel for different drain bias conditions. Fig. 19 plots the variation of surface channel potential in the silicon thin-film for different drain bias conditions for a 0.2 μm fully depleted SOI MOSFET. It is evident from the figure that the channel potential minima changes negligibly as the drain bias increases. Thus the incorporation of the DMG structure leads to an excellent immunity against DIBL in thin-film SOI devices. This is attributed to the step in channel potential profile due to the presence of the higher workfunction gate near the source. The negligible change in channel potential step at increasing drain bias due to the screening of the gate M1 is responsible reduction in channel length modulation (CLM). The model predictions correlate well with the simulation results obtained from MEDICI [79].

The drain side electric field pattern in the channel gives an indication of the magnitude of the hot-carrier effects. The electric field component in the *x*–direction, under the metal gates M1 and M2 is given as

$$E_1(x) = \frac{d\phi_1(x,y)}{dx}\bigg|_{y=0} = \frac{d\phi_{S1}(x)}{dx} = A\lambda_1\exp(\lambda_1 x) + B\lambda_2\exp(\lambda_2 x) \tag{9}$$

$$E_2(x) = \frac{d\phi_2(x,y)}{dx}\bigg|_{y=0} = \frac{d\phi_{S2}(x)}{dx} = C\lambda_1\exp(\lambda_1(x-L_1)) + D\lambda_2\exp(\lambda_2(x-L_1)) \tag{10}$$

Fig. 20 compares the surface electric distribution along the channel near the drain for SMG and DMG-SOI MOSFET's with a channel length $L = 0.4$ µm. As illustrated in the figure, the presence of lower workfunction gate on the drain side reduces the peak electric field considerably. This reduction in peak electric field consequently leads to a reduction in hot-carrier effects.

The electric field distribution along the channel also determines the electron transport velocity through the channel. Fig. 21 compares the surface electric field and the electron velocity profile in the thin-film in DMG SOI with a conventional SMG SOI MOSFET for a channel length of 0.3 µm at different drain bias conditions. The presence of the two different gate materials in a DMG SOI results in two peaks in the electric-field profile at the interface of the gate materials (one-third of the channel in this case). This leads to an enhancement of the electric field at the source side resulting in larger mean electron velocity when the electrons enter the channel from the source. It is observed that the step in the channel potential profile also forces the electric field to redistribute mostly at the drain side as the drain bias is increased.

Fig. 22 demonstrates the performance advantage of the DMG SOI MOSFET over its SMG counterpart. It is evident from the figure that incorporation of two different gate materials in a DMG structure leads to a simultaneous transconductance enhancement and drain conductance reduction. This highly desirable attribute is not easily achievable with other approaches to suppress SCE. The on-current reduction in DMG SOI MOSFET is because of an elevated threshold voltage as shown in Fig. 22 (b).

## 4. Conclusions

In sections 2 and 3 of this paper, the physical mechanisms responsible for short-channel effects (SCE) in SOI devices have been studied and recent attempts to alleviate the SCE have been reviewed. Specific strengths and weaknesses of the different approaches have been discussed. Engineering channel doping in a controlled way is a popular way but it becomes prohibitively difficult with extremely thin-films and scarce and randomly positioned dopant atoms, implying yield and reliability problems. On the other hand, very thin buried oxides ($t_b \sim 100$ nm) are needed to avoid coupling, which trades-off with junction capacitance considerations. Multiple gate SOIs like the Double-Gate (DG) SOI offer good immunity against SCE but there are difficulties to integrate them in the current CMOS fabrication technology.

*Dual-Material Gate* (DMG) SOI MOSFETs promise simultaneous suppression of SCE and enhancement of average carrier velocity in the channel. The efficacy of the DMG structure in suppressing SCE is assessed using a physics based two-dimensional analytical model of surface potential in the thin-film in section 4. Numerical simulation studies further demonstrate the simultaneous transconductance enhancement and output conductance reduction easily achievable by way of "gate-material engineering" in a thin-film DMG SOI MOSFET.

A major concern towards integrating DMG structure in the present SOI technology may arise from the fabrication viewpoint. But, Zhou [80] showed two alternative procedures to fabricate dual material gate structure in bulk CMOS technology with the addition of a single mask. Furthermore, the DMG structure may also be employed as an LDD spacer by adding a layer of material with different workfunction to both sides of the gate. With the aggressive scaling of CMOS processing technology the benefits of the excellent immunity against SCE's and simultaneous transconductance enhancement and output conductance reduction offered by the DMG SOI MOSFET position them as lucrative alternatives over the conventional SOI.

**Figure Captions**

Figure 1      Surface potential variation along the position in channel for 0.1 V and 1.5 V drain voltages (linear and saturated case).

Figure 2      Three mechanisms determining SCE in SOI MOSFETs [31].

Figure 3      Comparison of schematic energy band diagrams near the bottom of the body between the long and short-channel fully depleted (FD) nMOSFET's [31].

Figure 4      Effects of the three mechanisms on threshold voltage dependence on gate length [31].

Figure 5      Short channel effect in a FD SOI NMOS device with front gate oxide of 9.2 nm, buried oxide of 400 nm, thin-film of 80 nm, with back gate bias of 0 and –5 V [34].

Figure 6      Threshold voltage roll-off of FD SOI NMOS device with a front gate oxide of 4.5 nm and various thin-film thicknesses [35].

Figure 7      Threshold voltage shift versus thin-film thickness for various channel doping densities, biased at (a) $V_{DS} = 0.05$ V, and (b) 1.5 V [36].

Figure 8      Threshold voltage versus channel length of an SOI NMOS device with front gate oxide of 6 nm and a thin-film of 100 nm, and buried oxide of 100 nm and 400 nm [37].

Figure 9      Electric field lines from the drain [38].

Figure 10      Comparison of device structures for (a) a conventional MOS and (b) a raised source/drain thin-body transistor. Thin-body device structure can effectively suppress sub-surface leakage current [42].

Figure 11      Threshold voltage versus channel length of an FD SOI NMOS device using polysilicon and tantalum gates [49].

Figure 12      Threshold voltage roll-off due to DIBL and CS versus buried oxide permittivity [50].

Figure 13      Graded channel SOI MOSFET [51].

Figure 14      Cross-section of a single halo (SH) SOI nMOSFET [58].

Figure 15      Ground plane under (a) source and drain edge [63] or (b) channel region [64].

Figure 16      Double-gate, triple-gate, gate all around (GAA) and Π-gate SOI MOSFETs [72].

Figure 17      $V_{TH}$ roll-off and DIBL in double, triple, quadruple and Π-gate SOI MOSFETs. Device width and thickness = 30 nm [72].

Figure 18      Cross-sectional view of an n-channel fully depleted DMG-SOI MOSFET.

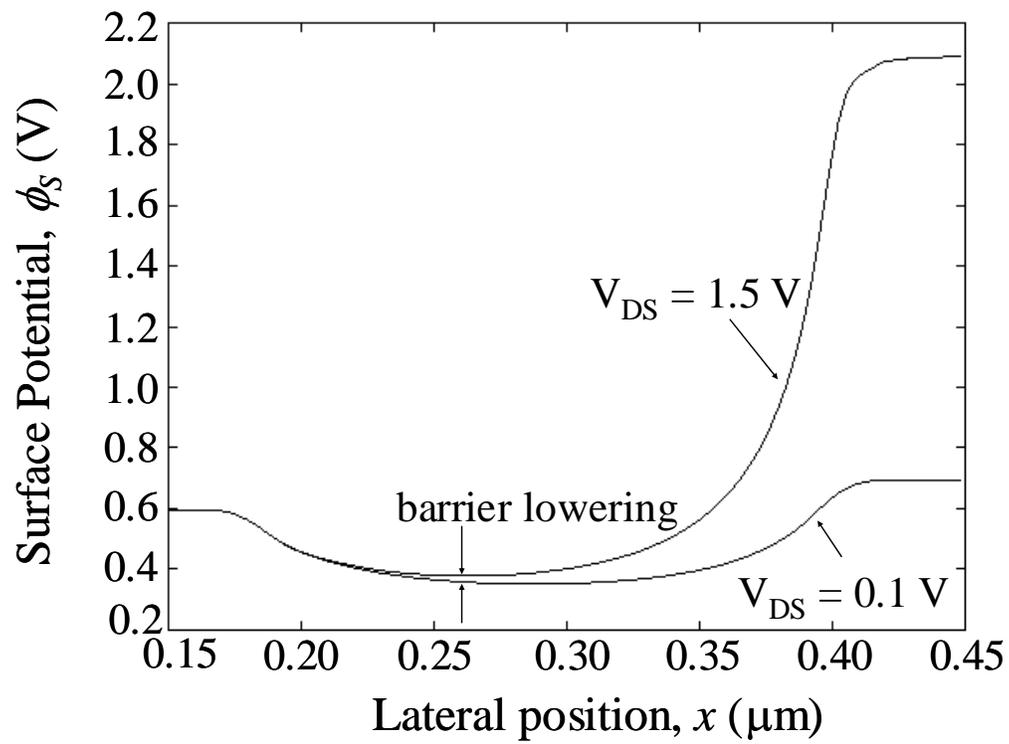

Figure 1.

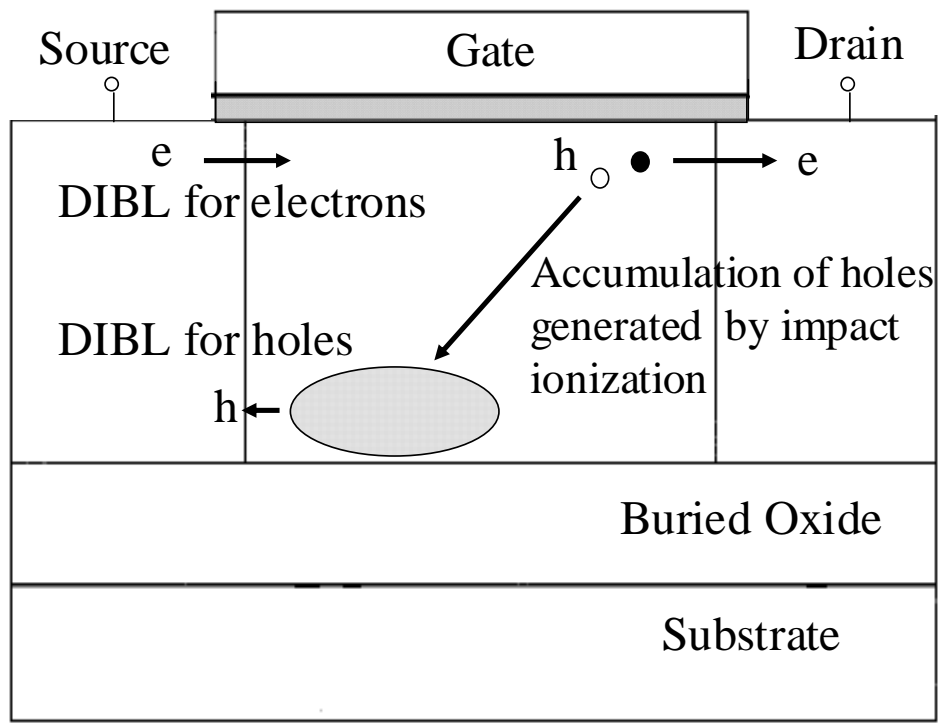

Figure 2.

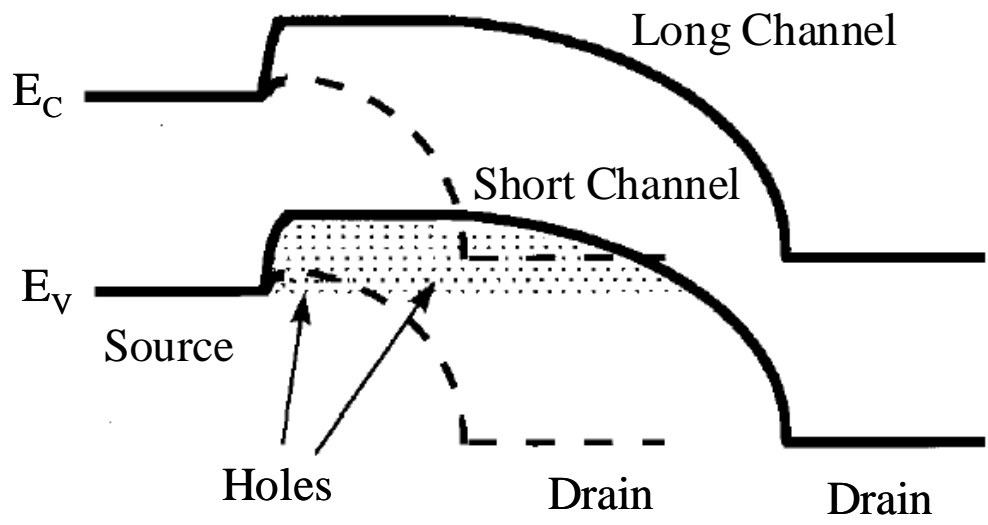

Figure 3.

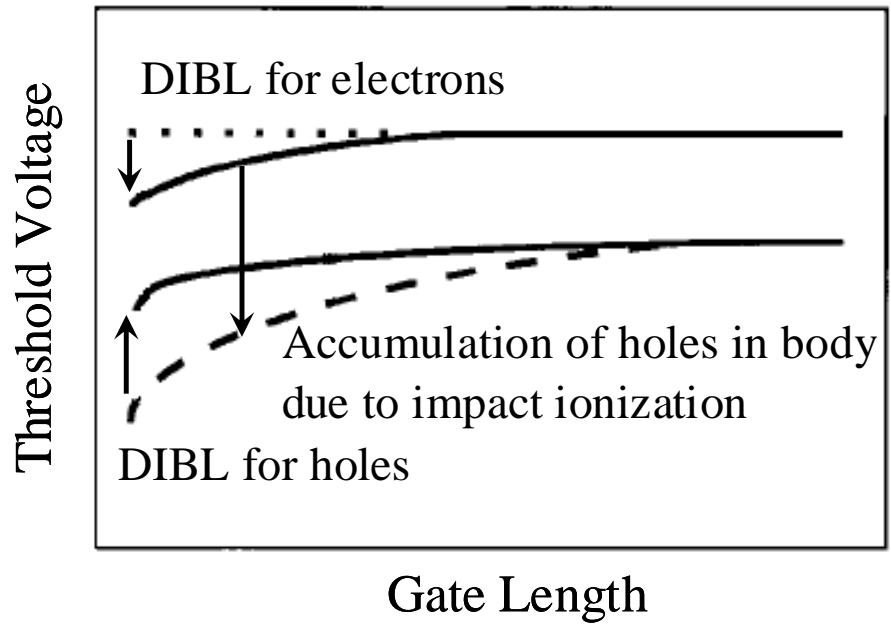

Figure 4.

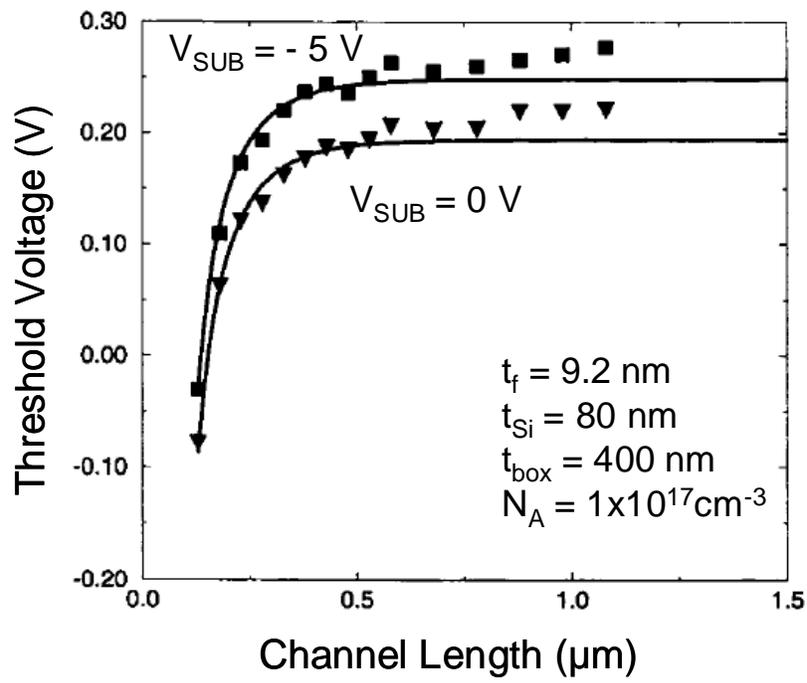

Figure 5.

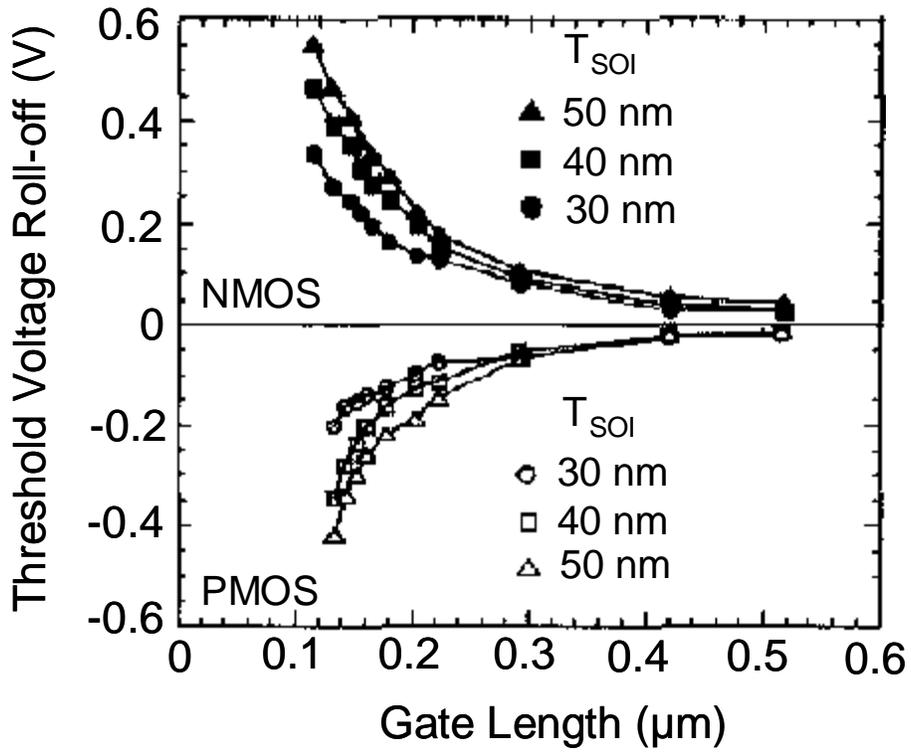

Figure 6.

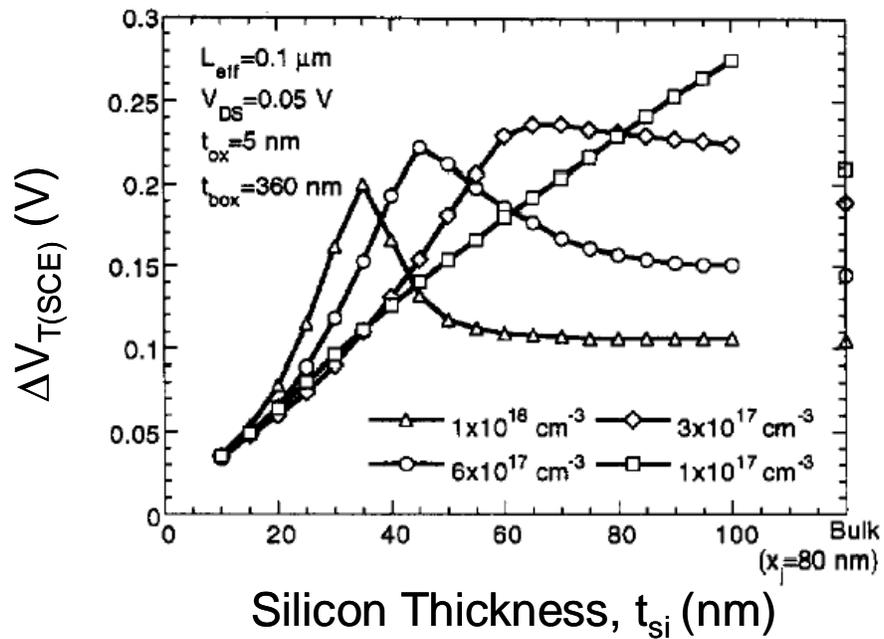

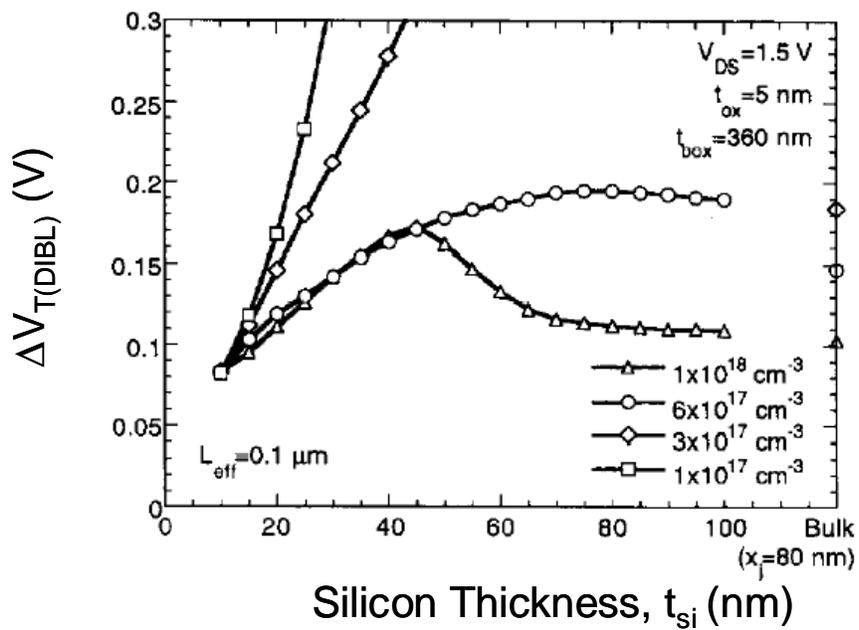

Figure 7.

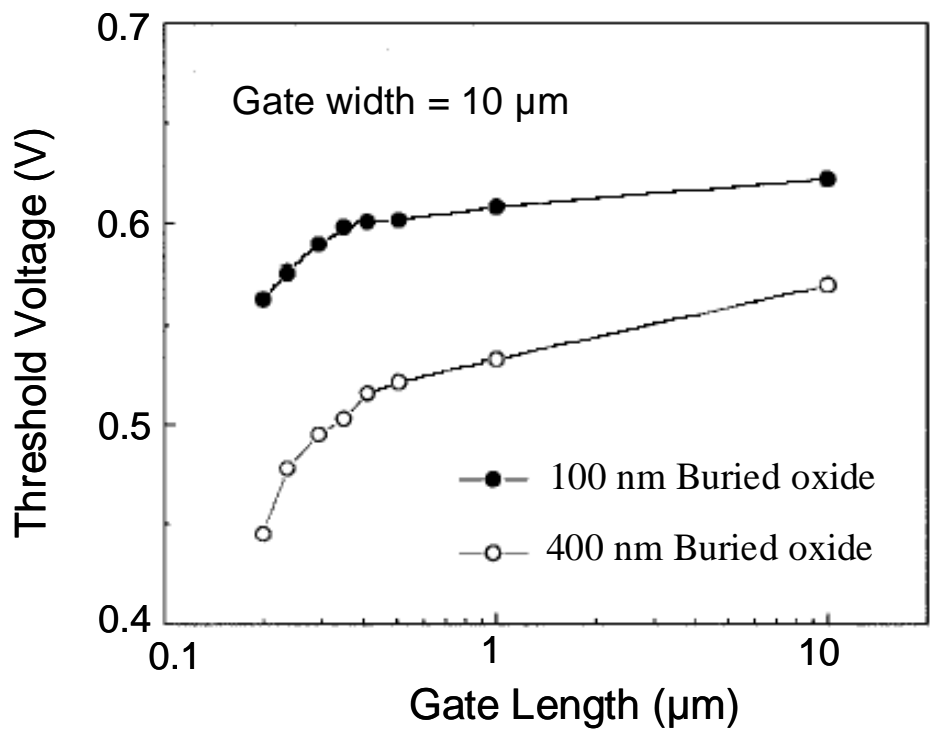

Figure 8.

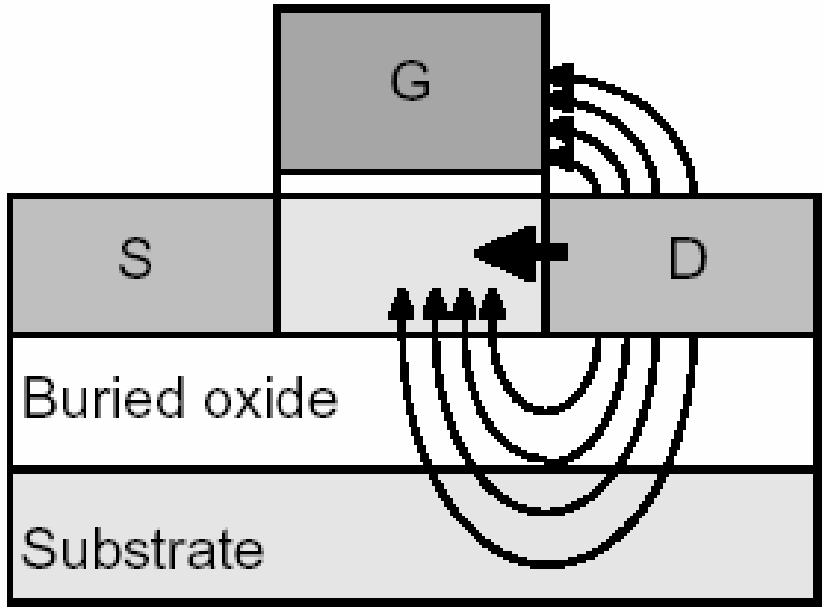

Figure 9.

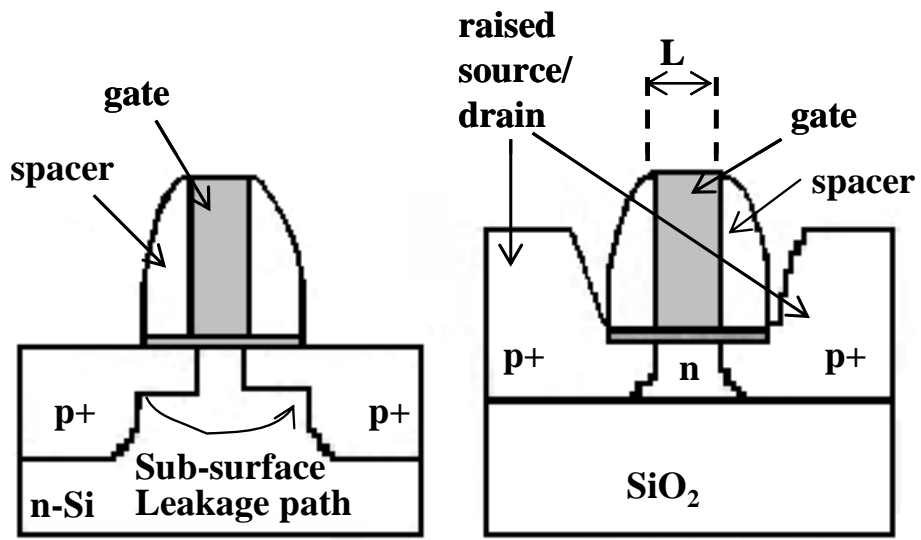

Figure 10.

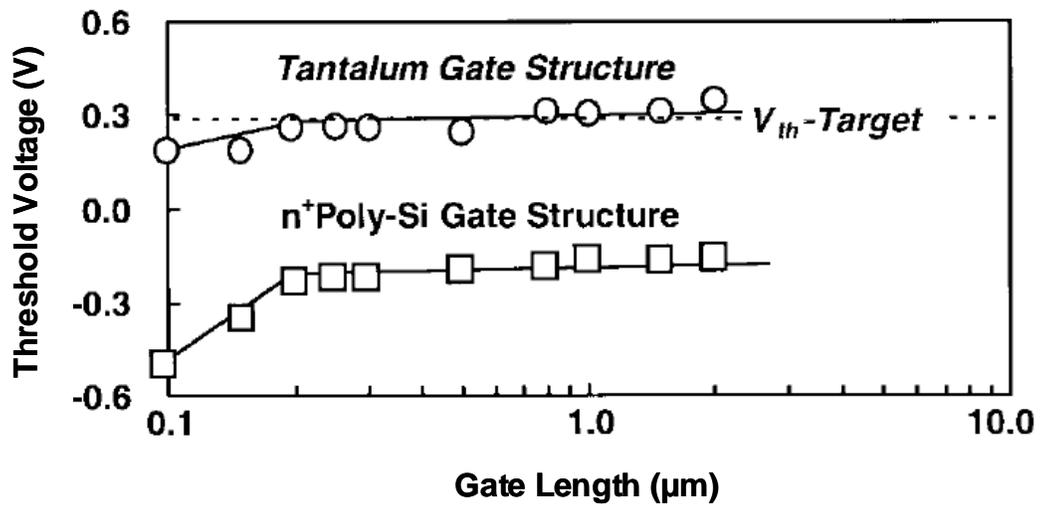

Figure 11.

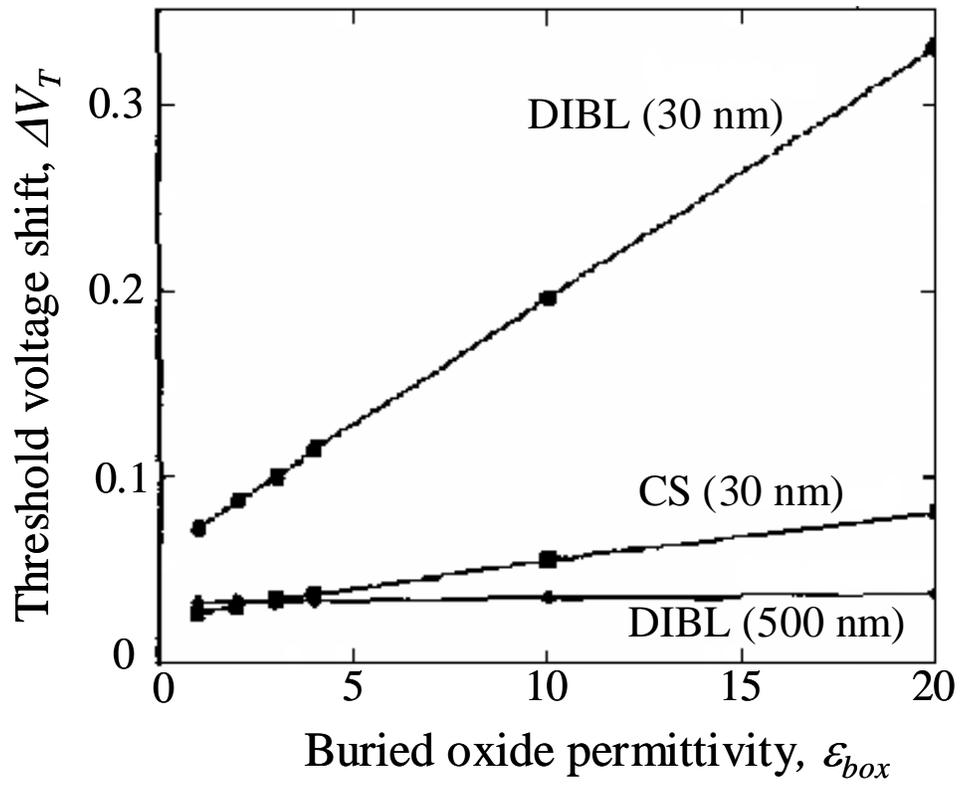

Figure 12

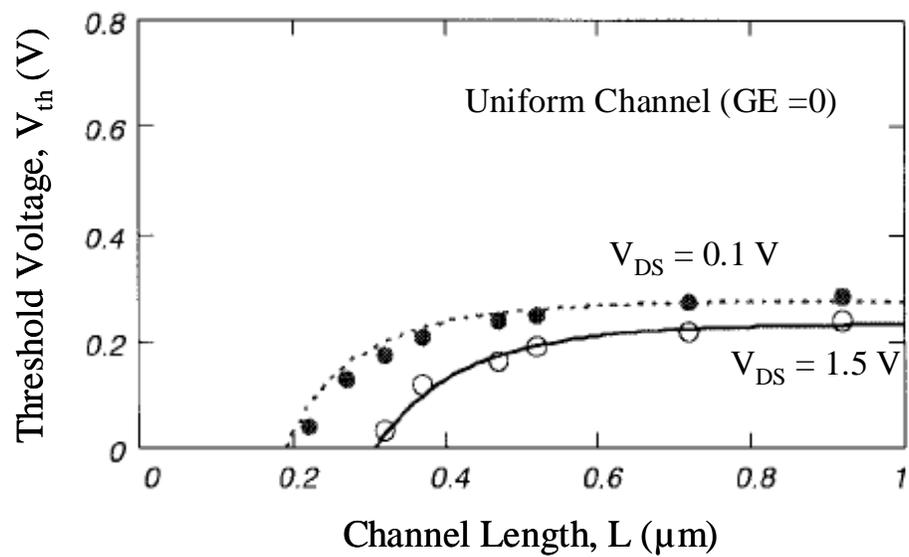
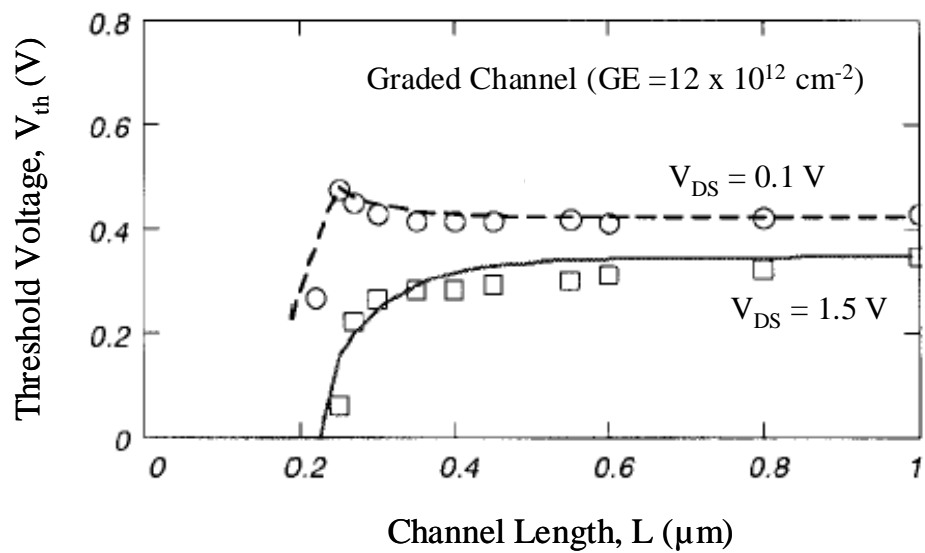
Figure 13.

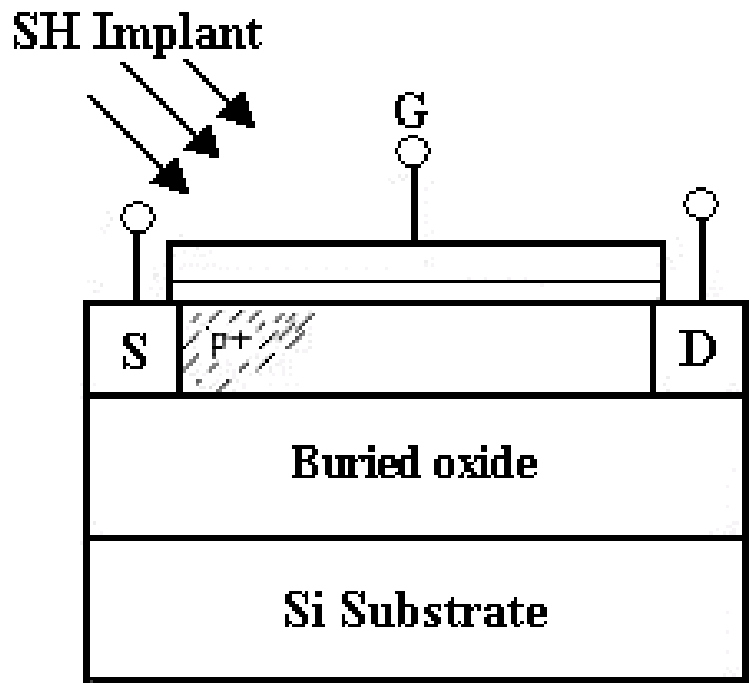

Figure 14

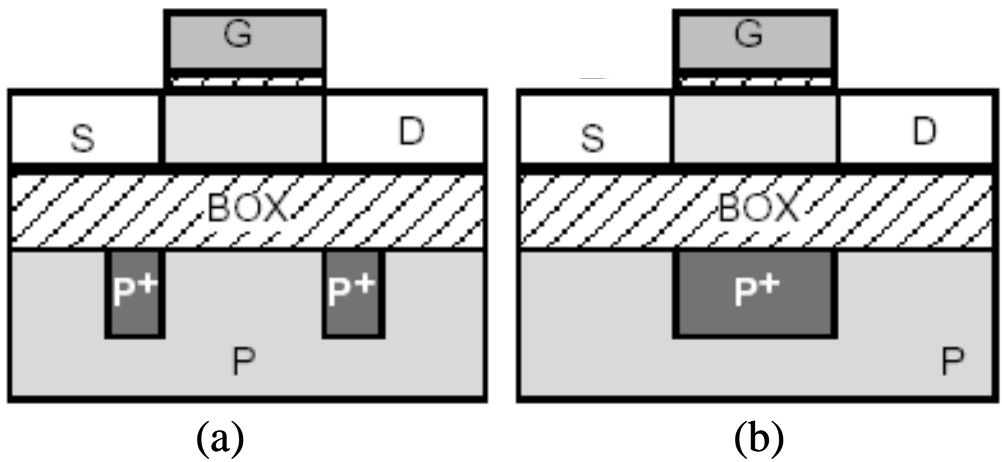

Figure 15.

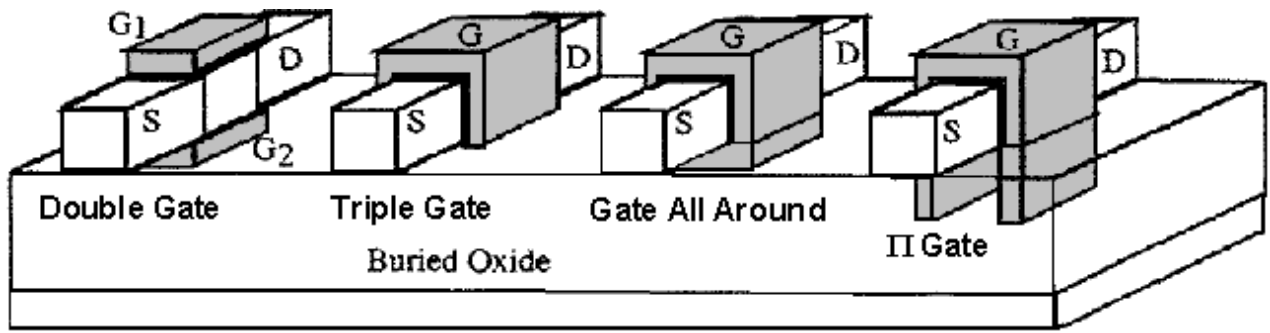

Figure 16.

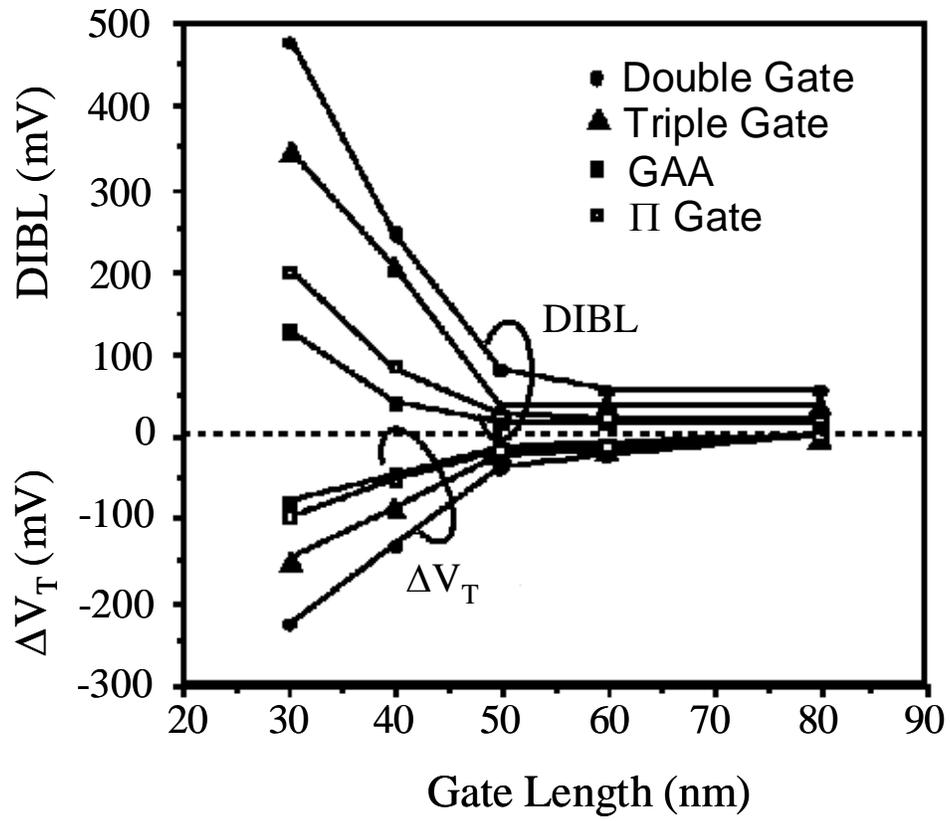

Figure 17.

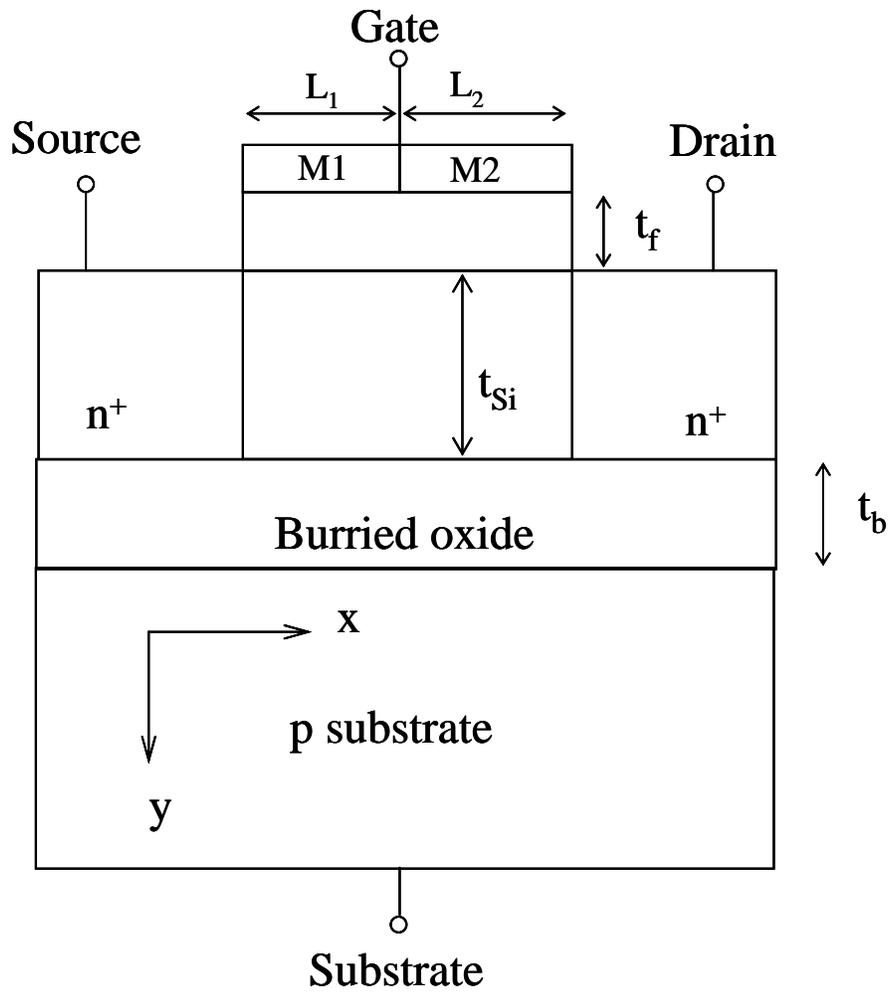

Figure 18.

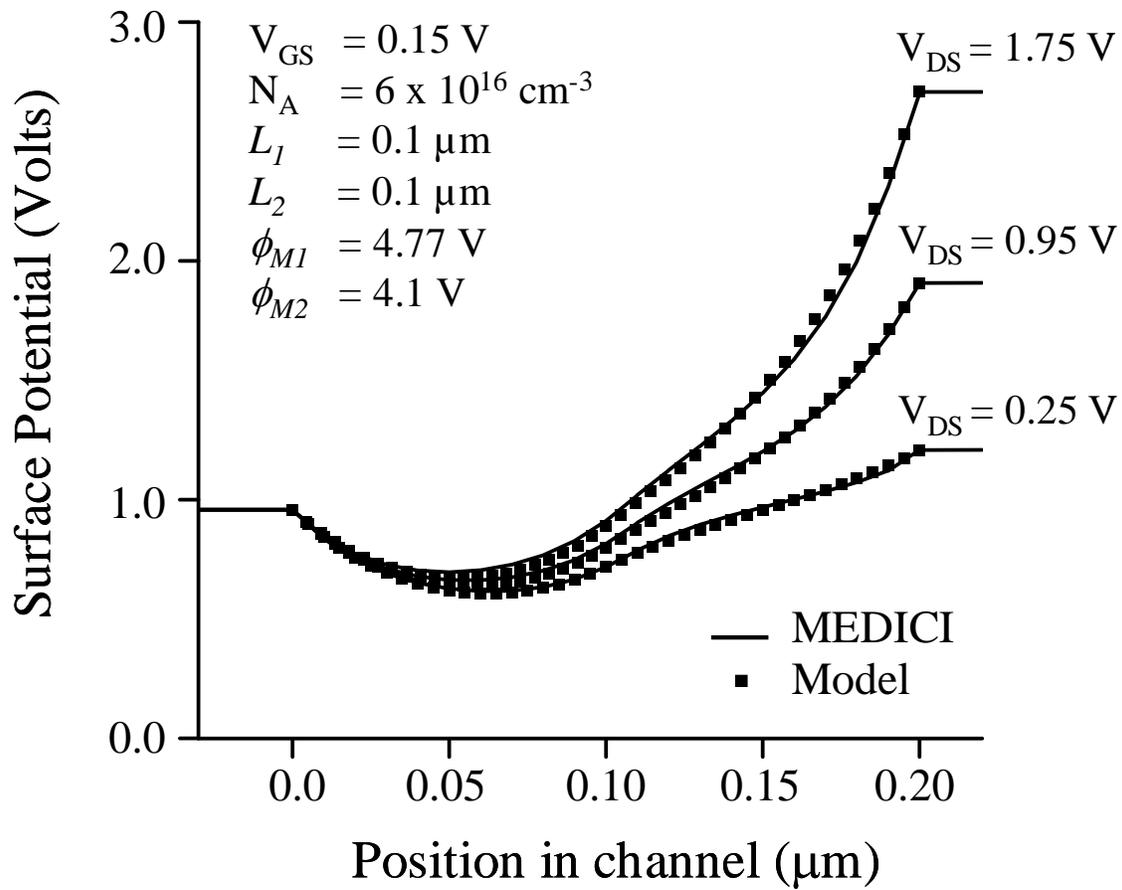

Figure 19.

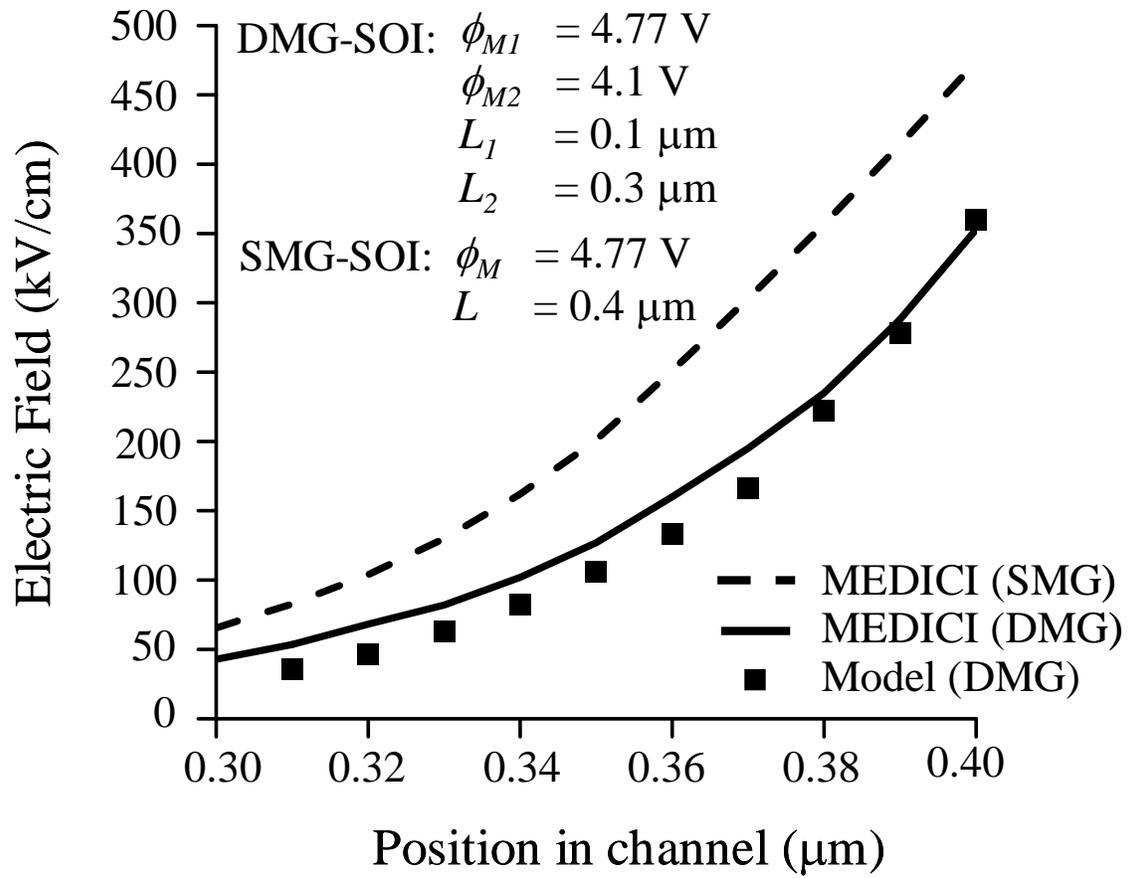

Figure 20.

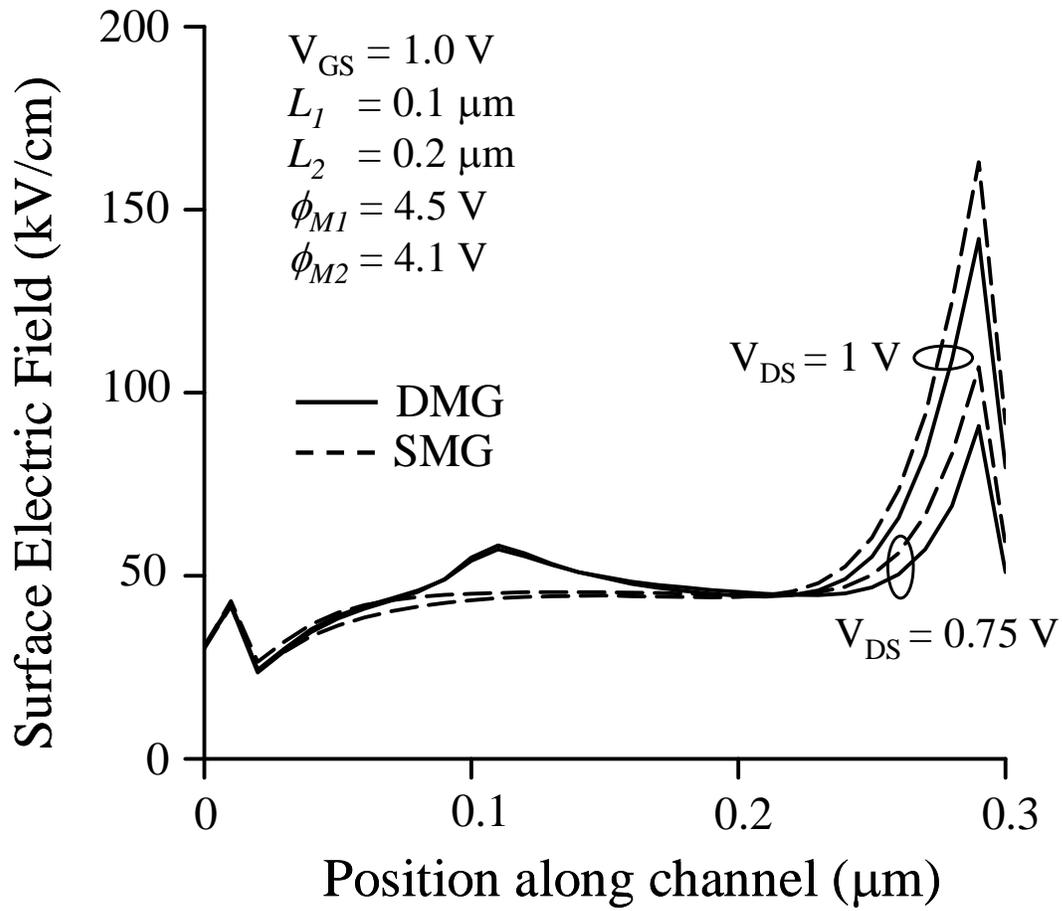

Figure 21(a).

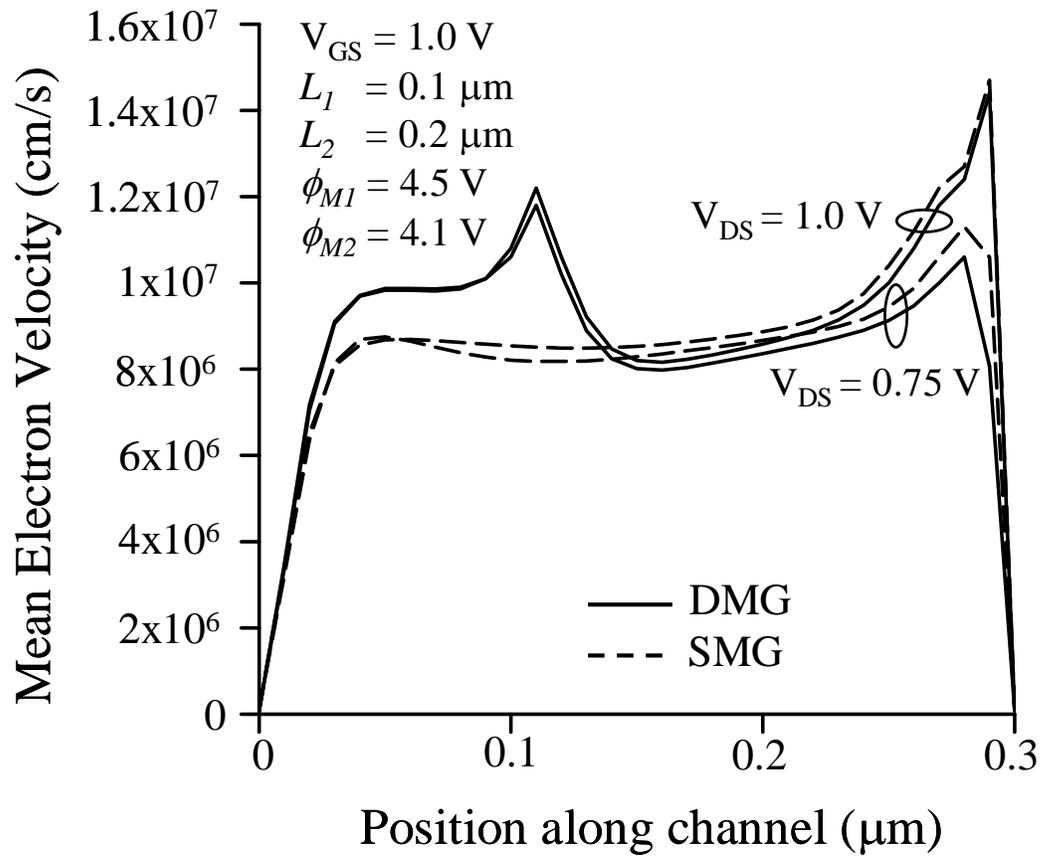

Figure 21(b).

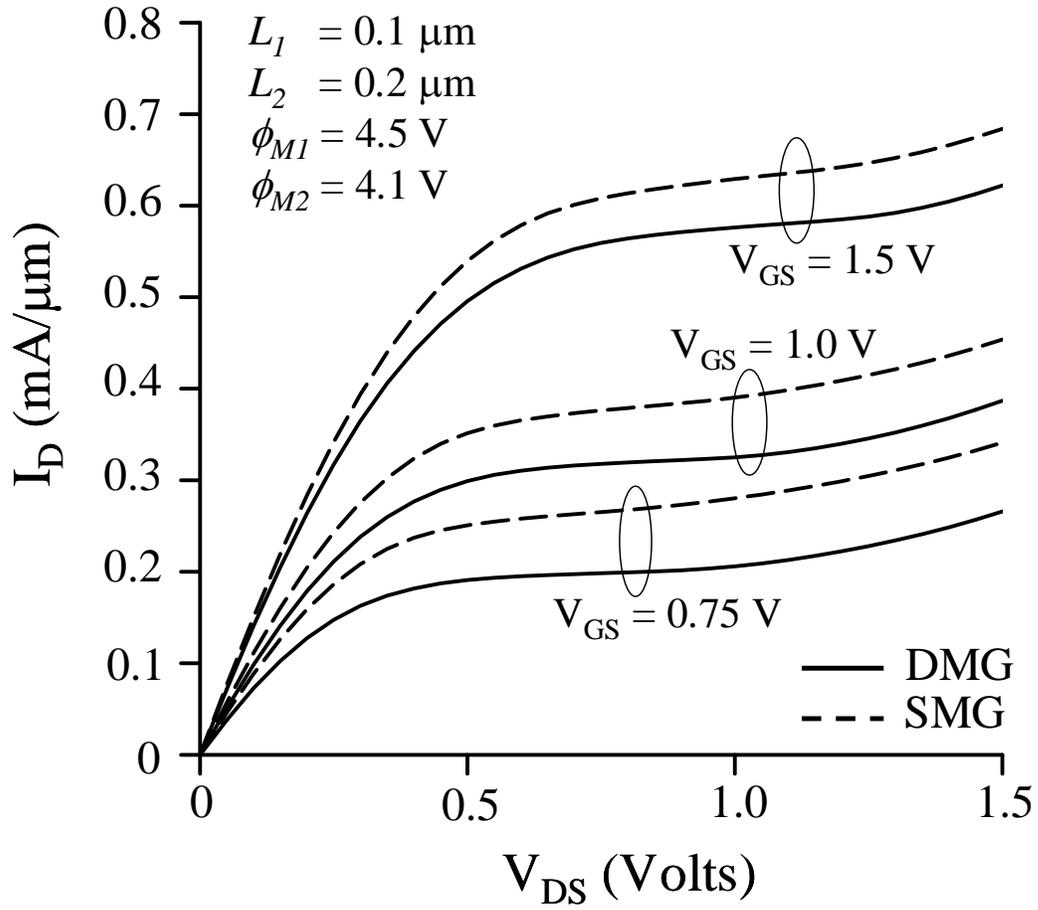

Figure 22(a).

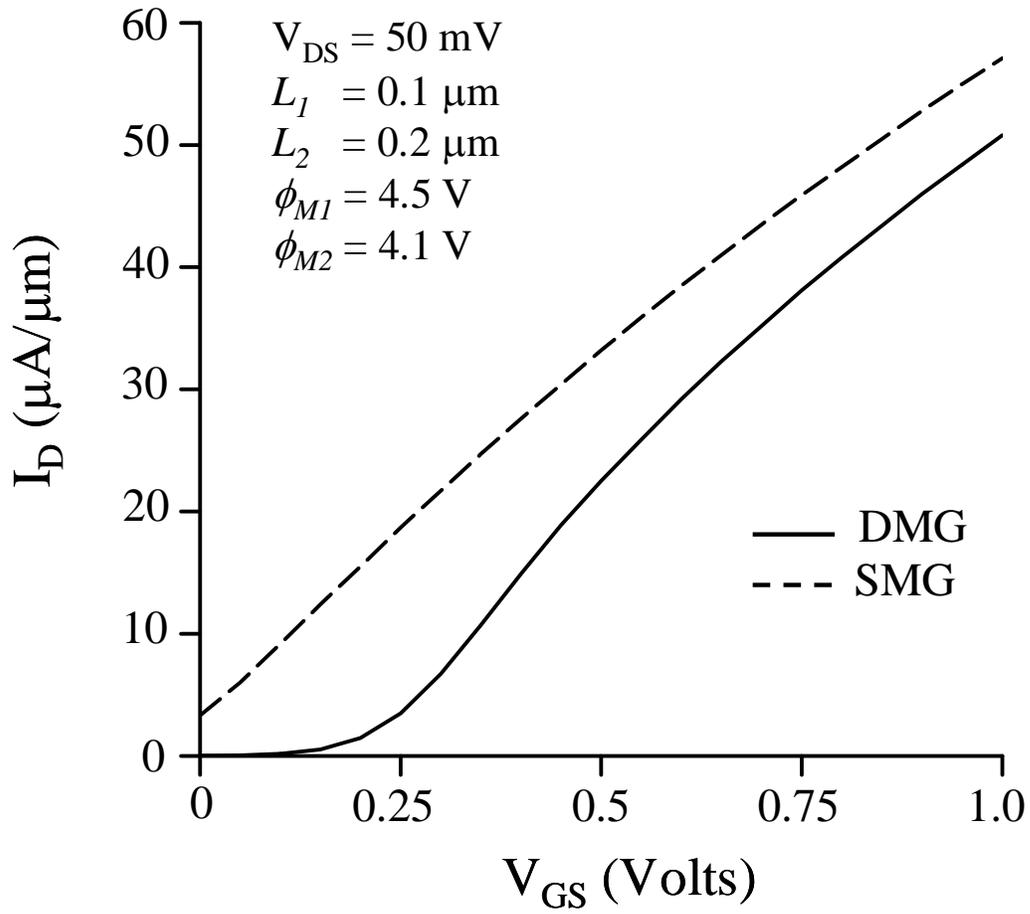

Figure 22(b).